\documentclass{iopart}
\usepackage{graphicx}
\usepackage{epsfig}
\begin{document}

\article[Improvement of Laser Cooling of Ions in a Penning
Trap]{PROCEEDINGS OF THE TCPFI 2002 CONFERENCE}{Improvement of
Laser Cooling of Ions in a Penning Trap by use of the Axialisation
Technique}

\author{H. F. Powell, S. R. de Echaniz, E. S. Phillips, D. M. Segal, and
R. C. Thompson}

\address{Quantum Optics and Laser Science Group,
Blackett Laboratory, Imperial College, Prince Consort Road, London
SW7 2BW, U.K.}

\begin{abstract}
We report a study of the axialisation and laser cooling of single ions and small clouds of ions in a Penning trap. A weak radiofrequency signal applied to a segmented ring electrode couples the magnetron motion to the cyclotron motion, which results in improved laser cooling of the magnetron motion. This allows us to approach the trapping conditions of a Paul trap, but without any micromotion. Using an ICCD camera we show that the motion of a single ion can be confined to dimensions of the order of 20 $\mu$m or less. We have measured increased magnetron cooling rates for clouds of a few ions, using an rf-photon correlation technique. For certain laser cooling conditions, the magnetron motion of the center of mass of the cloud grows and stabilises at a large value. This results in the ions orbiting the center of the trap together in a small cloud, as confirmed by photon-photon correlation measurements.
\end{abstract}

\pacs{32.80.Pj, 03.67.Lx, 41.20.-q, 42.50.Vk}

\submitto{\jpb}

\section{Introduction}

Trapped ions are currently being studied extensively with regard to their possible use in the realisation of a prototype quantum computer \cite{steane}. Laser-cooled trapped ions offer distinct advantages over many other proposed quantum technologies for this purpose. As a result, much work in the area of quantum information science has focused on the use of miniature radiofrequency (Paul) traps (for single ions) or linear radiofrequency traps (which allow the preparation of stationary strings of ions). In both cases, strong confinement to the Lamb-Dicke regime may be obtained (e.g. \cite{roos, nist}). Unfortunately, the low-temperature limit of the Paul trap can be affected by the micromotion of the trapped ions. This arises as a result of the applied radiofrequency trapping field and cannot be completely avoided, although the micromotion is generally minimised by the application of potentials to compensation electrodes. Decoherence rates may be limited by the presence of the micromotion or by the small size of the trap electrodes, which allow image charge and patch potential effects to become significant \cite{devoe}.

Ions in a Penning trap have no micromotion because the applied electric and magnetic fields are both static. Furthermore, since the radial confinement is given by the magnetic field, strong confinement does not require small electrodes as it does in the case of radiofrequency traps. However, the nature of the motion in the radial plane of the Penning trap complicates the laser cooling process (e.g. \cite{papa}). In the Penning trap, the radial motion (perpendicular to the magnetic field $B$) of an ion (mass $m$, charge $e$) is described by a combination of two circular orbits: a fast cyclotron motion at the modified cyclotron frequency $\omega_c^{\prime}$ and a slow magnetron motion at $\omega_m$. These frequencies 
are given by
\begin{equation}
	\omega_c^{\prime}=\omega_c/2 +\omega_1
\end{equation}
\begin{equation}
	\omega_m=\omega_c/2-\omega_1
\end{equation}
where
\begin{equation}
	\omega_1^2=\omega_c^2/4-\omega_z^2/2.
\end{equation}
$\omega_c=eB/m$ is the true cyclotron frequency and $\omega_z=\sqrt{4eV/mR^2}$ is the axial oscillation frequency (all formulae for frequencies have been quoted in angular frequency units). Here $V$ is the applied voltage and $R^2=r_0^2+2z_0^2$ is a parameter determined by the diameter of the ring electrode ($2r_0$) and the separation of the endcaps (2$z_0$).

It can be shown that it is difficult to decrease simultaneously the amplitudes of both the magnetron and cyclotron motions with conventional Doppler laser cooling \cite{itano}. This is because the magnetron motion is unstable: its total energy is negative. Therefore energy must be supplied in order to reduce the magnetron radius, which can be achieved only when the laser is placed at the point where the magnetron rotation moves the ions in the same direction as the laser photons. Even so, the cooling of the cyclotron motion is considerably more efficient than that of the magnetron motion. As a result, the final amplitude of the magnetron motion for a laser-cooled ion in a Penning trap is not as small as might be expected from the standard Doppler temperature limit, limiting the degree of localisation of the ion that can be achieved by laser cooling. This is in contrast to the radiofrequency trap, where very tight localisation may be achieved. In some cases, even cooling to the Lamb-Dicke regime is possible with Doppler cooling, without the need for more advanced laser cooling techniques \cite{gscooling}.

For large numbers of ions in a Penning trap, the ions form into a relatively dense cloud with a size limited by the effects of space charge. This cloud is effectively a one-component non-neutral plasma and the dynamics are best treated in terms of plasma parameters. Under these conditions, the plasma has a well-defined size and shape and it rotates as a whole at a rate
which depends on its density (this rotation is analogous to the magnetron motion of a single particle). Laser cooling can be used to cool and compress the cloud, but this process is limited by the dynamics of the trap \cite{perp}. However, the situation can be improved by imparting an external torque to the plasma. One way of doing this is to use the \lq\lq rotating wall" technique. This increases the plasma density by locking the rotation of the plasma to an external drive field and increasing this rotation frequency to half the true cyclotron frequency, which corresponds to maximum plasma density \cite{wall}.

For the case of a single ion in a Penning trap, the motion is described using the single-particle parameters introduced above. This treatment also holds well for a few ions, when the cloud is not large enough to behave as a plasma. In our experiments we always studied
a small number of ions, so in the rest of this paper we use the single-particle treatment of the ion motion.

\section{Axialisation}

Axialisation is the process by which particles in a Penning trap are driven towards the central axis of the trap \cite{savard,schweik}. It is also sometimes referred to as azimuthal quadrupole excitation or as sideband excitation of the cyclotron motion. It occurs when the efficient cooling of the cyclotron motion is effectively extended to the magnetron motion by means of a coupling between the two. This results in a reduction in the amplitudes of both motions. Axialisation has been used in conjunction with buffer gas cooling to increase storage times in Penning traps and to increase the density of ion clouds \cite{konig}. A related technique is also used in precision mass measurements \cite{pritchard}. Combined with laser cooling, one expects very low effective temperatures to be reached using axialisation. 

The process of axialisation is driven by a radial quadrupole field at $\omega_c=\omega_c^{\prime}+\omega_m$ generated by a ring electrode split into four quadrants. This couples the magnetron and cyclotron motions. The quadrupole geometry is needed in order to drive the correct combination of motional frequencies; other geometries select different combinations \cite{marshall}. In the frame which rotates at $\omega_c/2$, in which the magnetic field is effectively eliminated by the rotation, the applied field generates an additional {\it static} quadrupole potential which couples the clockwise and anticlockwise motions in this frame. These are both at angular frequency $\omega_1$ and correspond to the cyclotron and magnetron motions in the laboratory frame. In the absence of additional cooling, the coupling gives a slow cycling of the motion between the cyclotron and magnetron motions. This effect has been employed in precision mass measurements \cite{massmeas}.

\begin{figure}
  \begin{center}
  	\includegraphics[width=8cm]{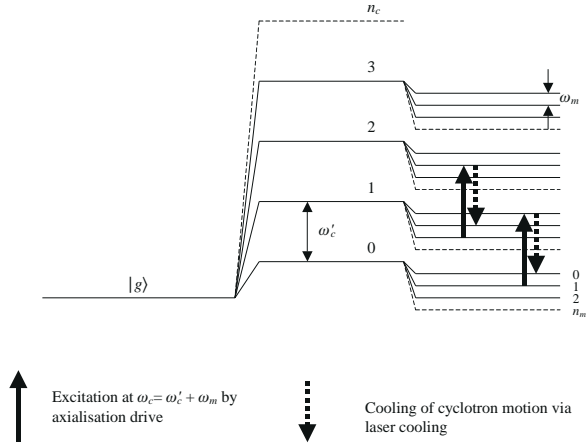}
	\end{center}
  \caption{\label{figure1}Motional energy levels of an ion in the Penning trap (not to scale). Note that the cyclotron energy is positive but that of the magnetron motion is negative. The solid bold upward arrow shows transitions driven by the axialisation drive at $\omega_c=\omega_c^{\prime}+\omega_m$, and the downward dashed arrow represents the effect of a cycle of laser cooling reducing the quantum number of the cyclotron motion. In our experiments, $\omega_c^{\prime}$ and $\omega_m$ are typically $2\pi\times$ 600 kHz and $2\pi\times$ 30 kHz respectively.}
\end{figure}

In a quantum mechanical picture, each excitation at the sum frequency can be thought of as causing the quantum number of the cyclotron motion to increase by 1 and that of the magnetron motion to {\it decrease} by 1. This reduces the magnetron radius $r_m$ and increases the cyclotron radius $r_c$. The excess energy in the cyclotron motion is then rapidly removed by the laser cooling. The reverse process also occurs but this is generally weaker (see below). 
An intuitive picture of the axialisation process is illustrated schematically in Figure \ref{figure1}, where the axialisation drive produces the transition $(n_c, n_m) \rightarrow 
(n_c+1, n_m-1)$ and then the laser cooling, via excitation to the excited state, effectively produces the transition $(n_c+1, n_m-1) \rightarrow (n_c, n_m-1)$. The overall change, for the particular sequence of events illustrated, is therefore to reduce the magnetron motion quantum number by 1. Note, however, that because the sidebands on the optical transition due to the ion motion are not resolved in our case, the actual change of $n_c$ in a particular optical excitation may be different from 1.

\begin{figure}
  \begin{center}
  	\includegraphics[width=8cm]{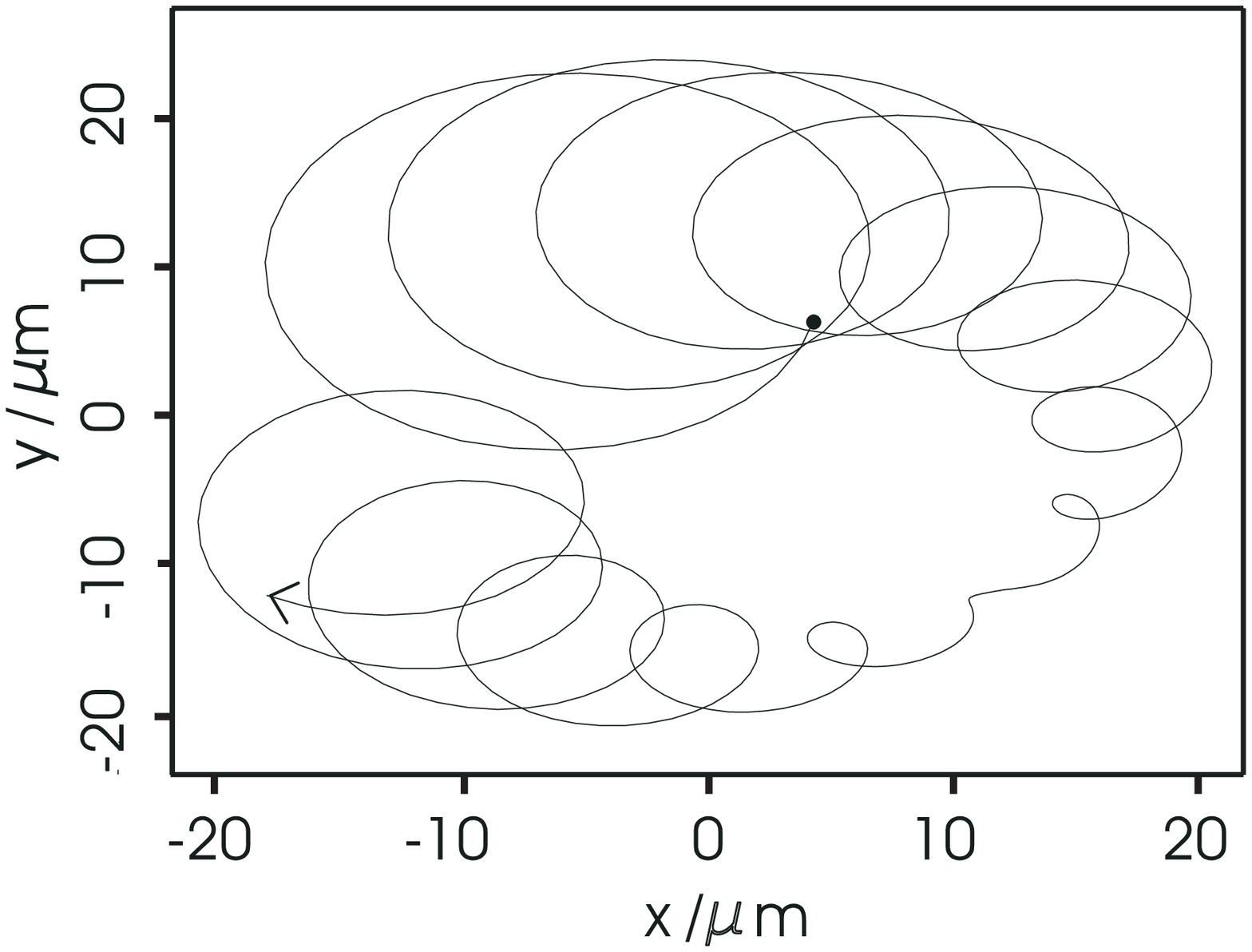}
  	\includegraphics[width=8cm]{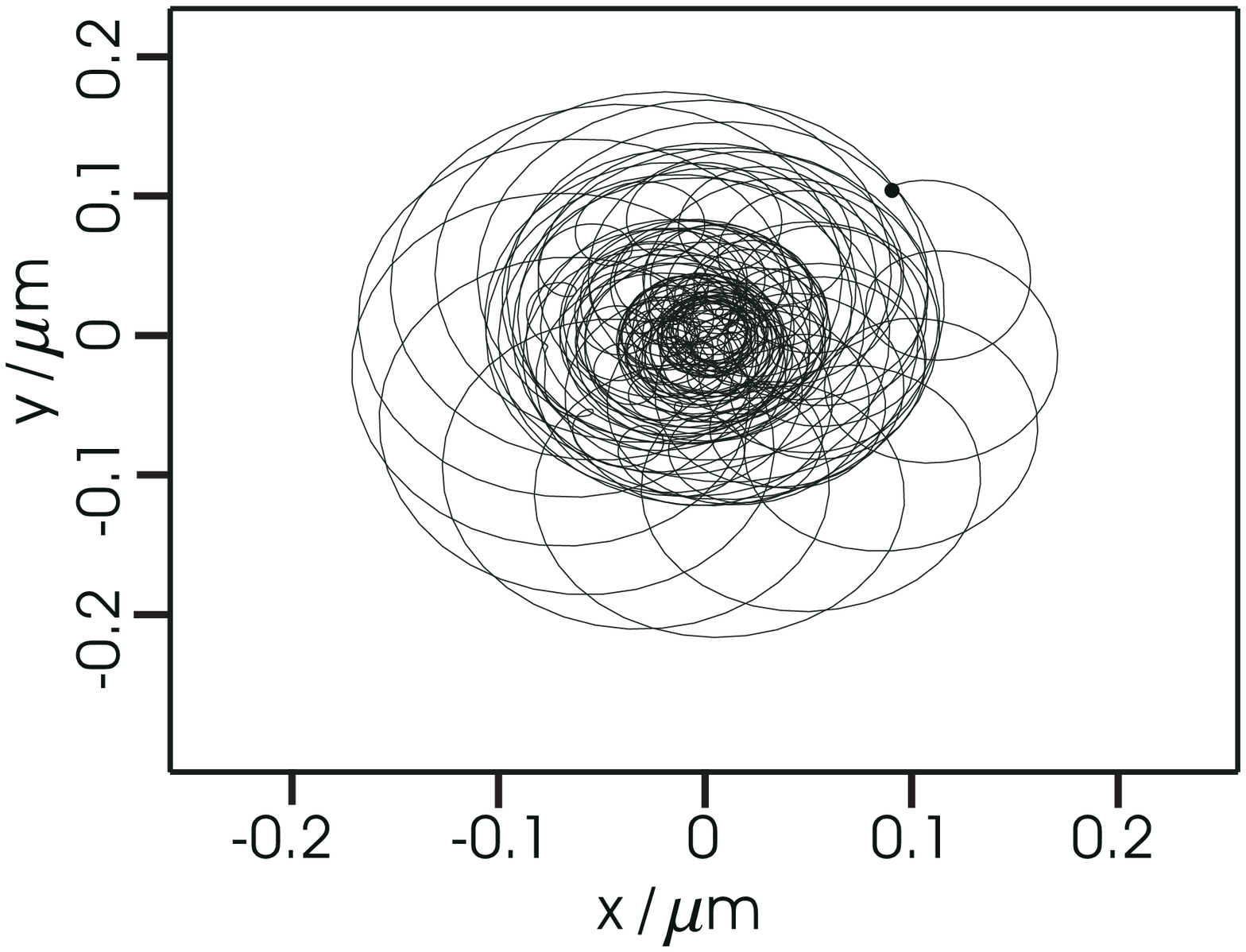}
  	\includegraphics[width=8cm]{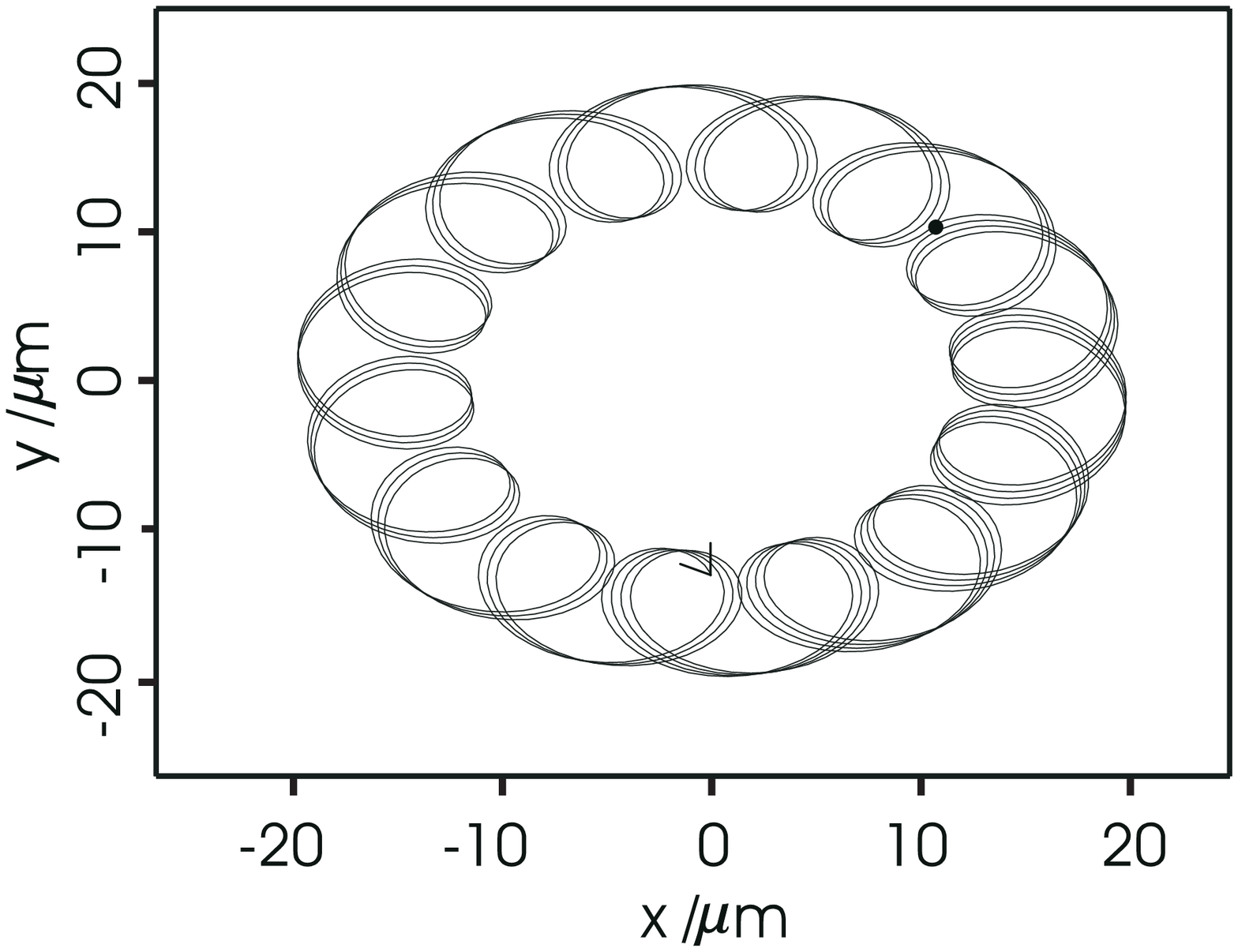}
  \end{center}
  \caption{\label{fig2}Simulation of ion motion in the presence of the axialisation drive: (a) with no cooling the ion cycles between cyclotron and magnetron motions; (b) with weak laser cooling the ion spirals towards the centre of the trap; (c) with strong laser cooling a stable orbit may be obtained.}
\end{figure}

The process can be described classically by the following equations:
\begin{equation}
	\mathrm{d}r_c/\mathrm{d}t = \delta\, r_m - \gamma_cr_c, 
\end{equation}
\begin{equation}
	\mathrm{d}r_m/\mathrm{d}t = - \delta\, r_c - \gamma_mr_m
\end{equation}
where $\gamma_c$ and $\gamma_m$ are the cooling rates for the two motions and $\delta$ is the coupling rate due to the rf axialisation field. Assuming that the diameter of the ion's motion is initially smaller than the laser beam, various solutions to the equations can be found: in 
Figure \ref{fig2} we show simulations of (a) oscillation between magnetron and cyclotron motions when the two damping rates are zero; (b) axialisation when $\delta^2 > - \gamma_c\gamma_m$; and (c) a stable orbit with finite radius when $\delta^2 = - \gamma_c\gamma_m$. This requires the cyclotron and magnetron damping rates to be of opposite sign: this situation arises in the Penning trap when the laser beam passes through the centre of the trap.

If the coupling rate $\delta$ is small and the damping rates are of opposite sign, then $\delta^2 < - \gamma_c\gamma_m$ and the orbit starts to expand. Because of the finite size of the laser beam, the orbit then takes the ion out of the laser beam for some of the time. The interaction with the laser therefore weakens, reducing the values of the damping rates. This process continues until the above equality in (c) is satisfied, resulting in a stable orbit whose size depends on the laser parameters. Note that with laser cooling the values of the damping coefficients can be varied by changing the laser frequency and position \cite{papa}. 
The axialisation process was previously demonstrated with buffer-gas cooling where the relative damping rates are fixed and do not vary with the amplitude of the motion. The appearance of the
stable orbit is therefore a feature unique to laser cooling.

\section{Demonstration of Axialisation}

Our set-up consists of a Penning trap with an internal diameter of 10 mm located between the pole pieces of an electromagnet generating a field of roughly 1 T \cite{linear}. The ring electrode is split into four segments for application of the quadrupole axialisation drive. The trapped magnesium ions are Doppler cooled with laser radiation tuned close to the resonance line at 280 nm. The natural linewidth of this transition is 43 MHz (giving a theoretical Doppler limit of 1 mK) so the motional sidebands are not resolved (all the ion oscillation frequencies are less than 1 MHz). The fluorescence is detected by a photomultiplier which feeds pulses to a multichannel scaler (MCS) (to measure the fluorescence rate) or to a time to amplitude converter (TAC) and multichannel analyser (MCA) (for photon correlation studies). Some of the light can be imaged with a magnification of $\sim 1$ onto an ICCD having a pixel size of 13 $\mu$m. If the fluorescence signal is monitored by the photo-multiplier, an increase is observed immediately when the axialisation drive is applied, as the size of the cloud is reduced and the ions are cooled more strongly \cite{powell}.

\begin{figure}
  \begin{center}
  	\includegraphics[width=8cm]{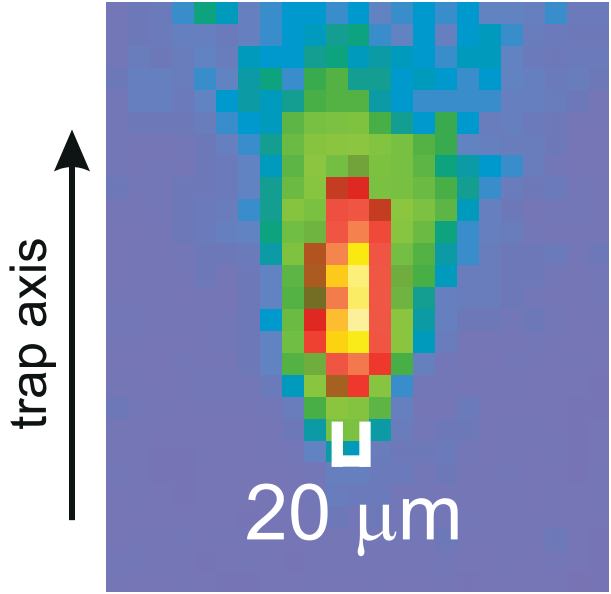}
	\end{center}
  \caption{\label{fig3}Image of a single ion with optimal axialisation and laser cooling.}
\end{figure}

\begin{figure}
  \begin{center}
  	\includegraphics[width=8cm]{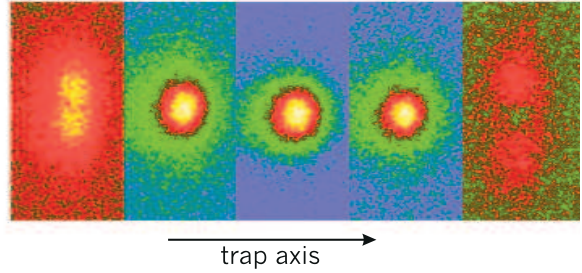}
	\end{center}
  \caption{\label{fig4}Images of a small axialised cloud with different axialisation drive frequencies close to resonance. The frequencies range from 621 to 631 kHz from left to right in 2 kHz steps; the fourth trace is at the true cyclotron frequency.}
\end{figure}

The ICCD images show an immediate reduction in the size of an ion cloud when the axialisation drive is turned on, showing that compression of the cloud has been achieved. It is then possible to optimise the conditions to improve the axialisation, which further reduces the size of the cloud. Figure \ref{fig3} shows an image of an optimised single ion. The measured size of the image is determined to be of the order of 20 $\mu$m, which is comparable to the finite resolution of the camera and imaging system. The equivalent temperature for the single ion is of order 10 mK \cite{powell}. This should be seen as an upper limit because of the finite resolution of the camera and imaging system. The observed shape of the cloud depends critically on the axialisation parameters, as can be seen in Figure \ref{fig4}, which shows the size and shape of an ion cloud for various detunings of the rf drive, close to the cyclotron frequency. 
This demonstrates that axialisation is most effective at the true cyclotron frequency.

\section{Measurement of Damping Rates}

\begin{figure}
  \begin{center}
  	\includegraphics[width=8cm]{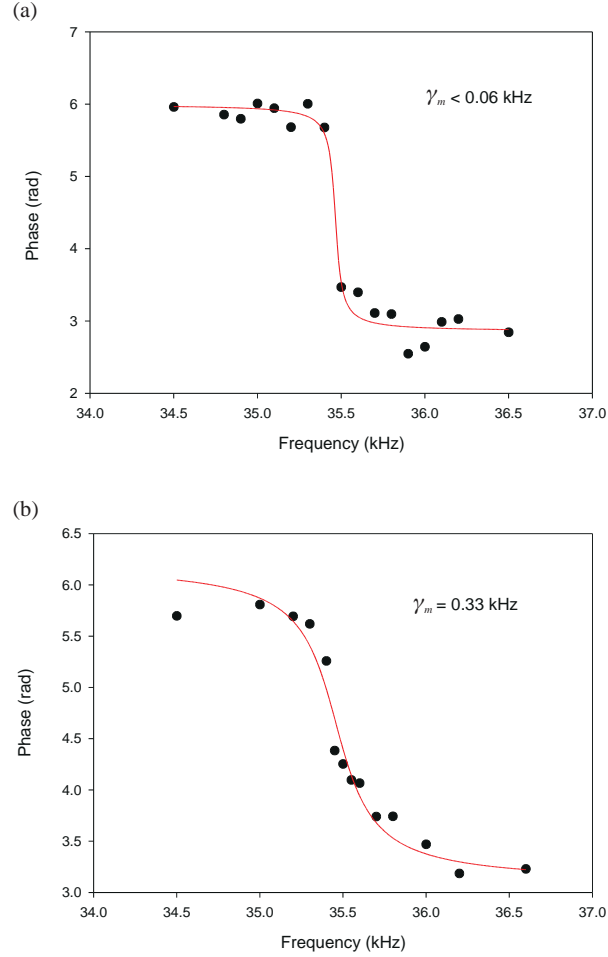}
	\end{center}
  \caption{\label{fig5}Plots of the phase of the driven magnetron motion for a small cloud of ions as a function of frequency of the drive close to the magnetron frequency. (a) with no axialisation present; (b) with axialisation present at an amplitude of 1.5 V.}
\end{figure}

We have measured the cooling rates for the motions of ions in a small cloud using an rf-photon correlation technique (see \cite{cooling}). By driving the ions with an additional, weak rf dipole field close to a motional resonance frequency, the phase difference between this dipole drive and the resulting motion can be found. Moving through the resonance, this phase difference changes by $\pi$ (as for any driven harmonic oscillator). The frequency width of this curve gives the cooling rate ($\gamma$) directly. Two examples (for the magnetron motion) are shown in Figure 5. In earlier experiments \cite{cooling} we measured this rate for 
both cyclotron and magnetron motions and found the magnetron cooling rate ($\gamma_m$) to be very low, as expected from calculations \cite{papa}. Measured damping rates for the magnetron
 motion were less than 100 Hz, whereas the cyclotron damping rate ($\gamma_c$) for magnesium was generally a few kHz, depending on the laser parameters and the number of ions. We have now 
measured the rate in the presence of axialisation (see Figure \ref{fig5}) and we find that the magnetron cooling is enhanced due to the axialisation, confirming that the strong cooling of the cyclotron motion is extended to the magnetron motion by this technique. Without the axialisation we find a cooling rate lower than 60 Hz (similar to our earlier measurements), 
but this increases to over 300 Hz after optimisation of the axialisation. The increase in cooling rate varies approximately linearly with the drive amplitude. These measurements
were not performed on a single ion, and the rates are expected to be significantly higher for single particles.

\section{Stable Orbits}

\begin{figure}
	\begin{center}
		\includegraphics[width=8cm]{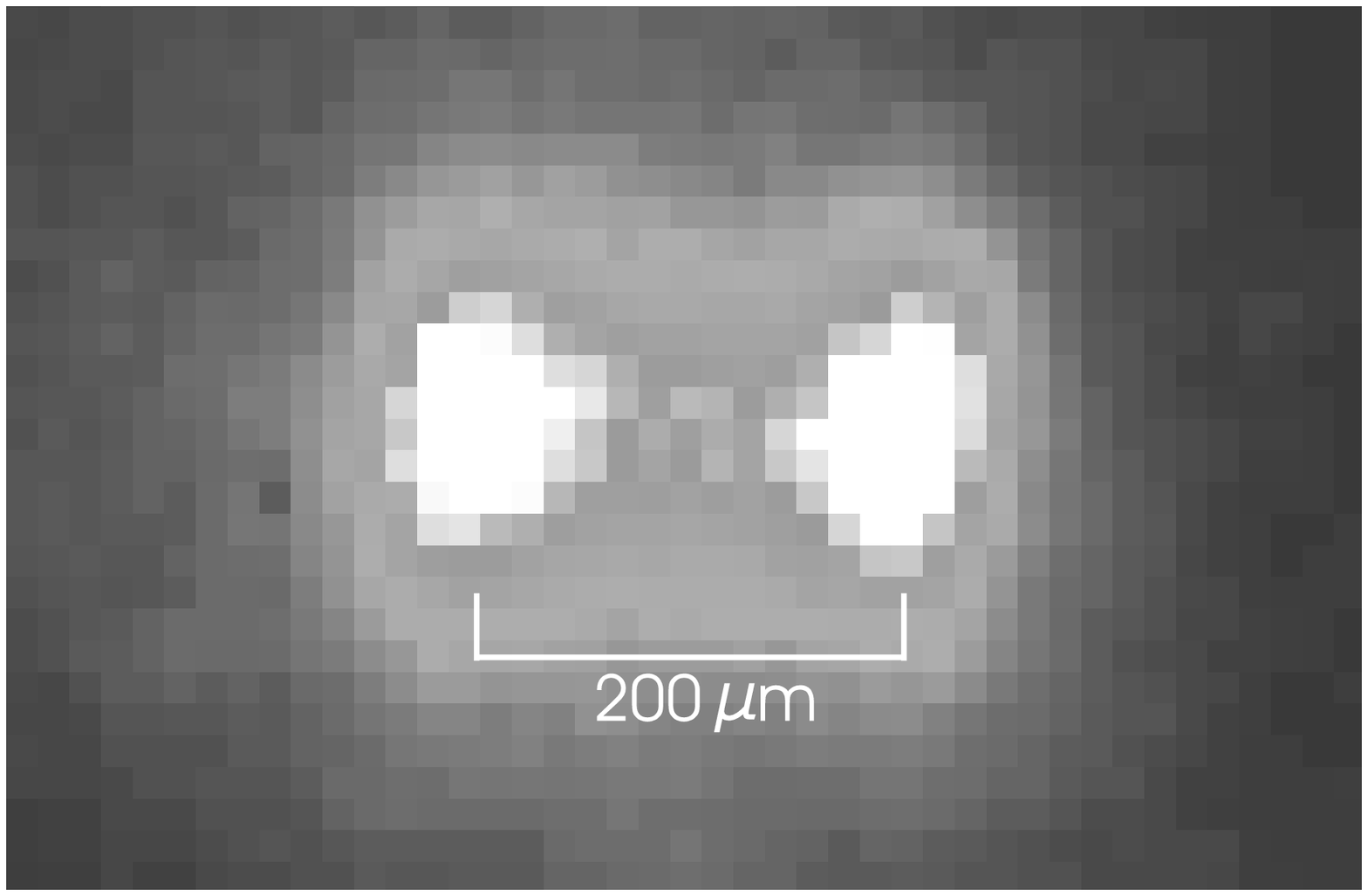}
		\includegraphics[width=9cm]{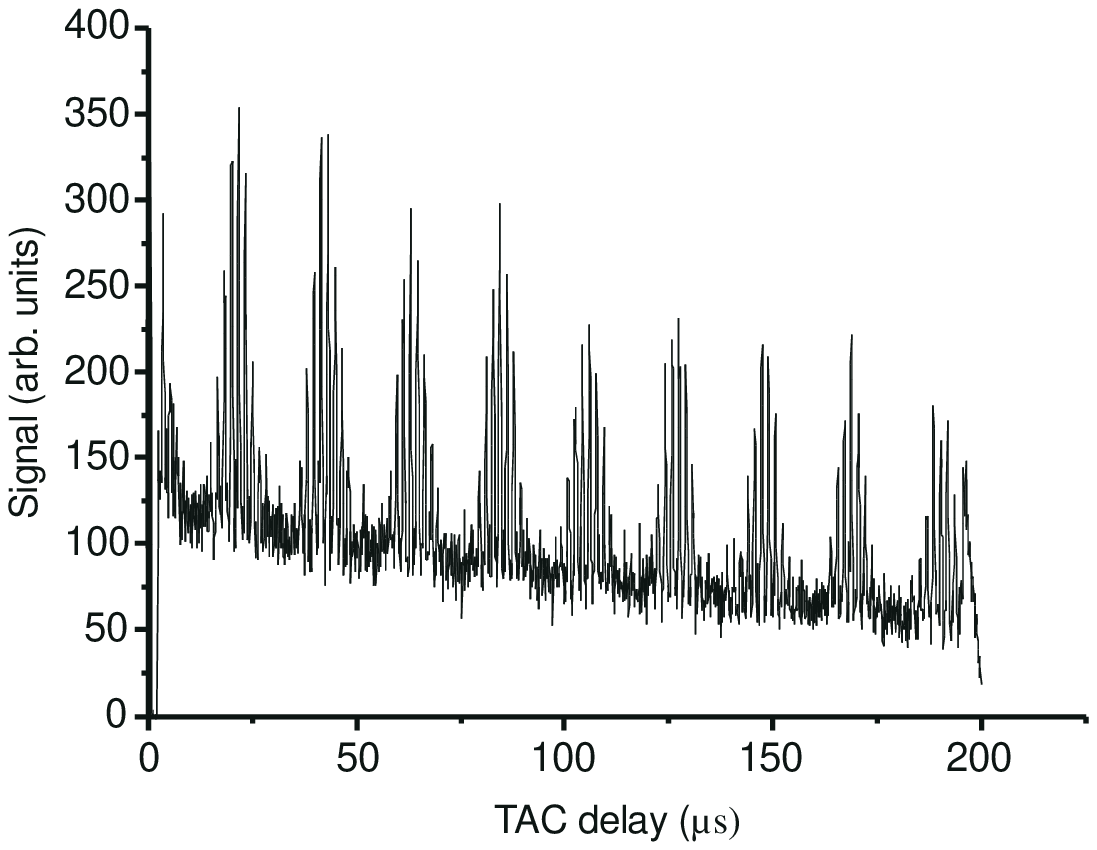}
		\includegraphics[width=10cm]{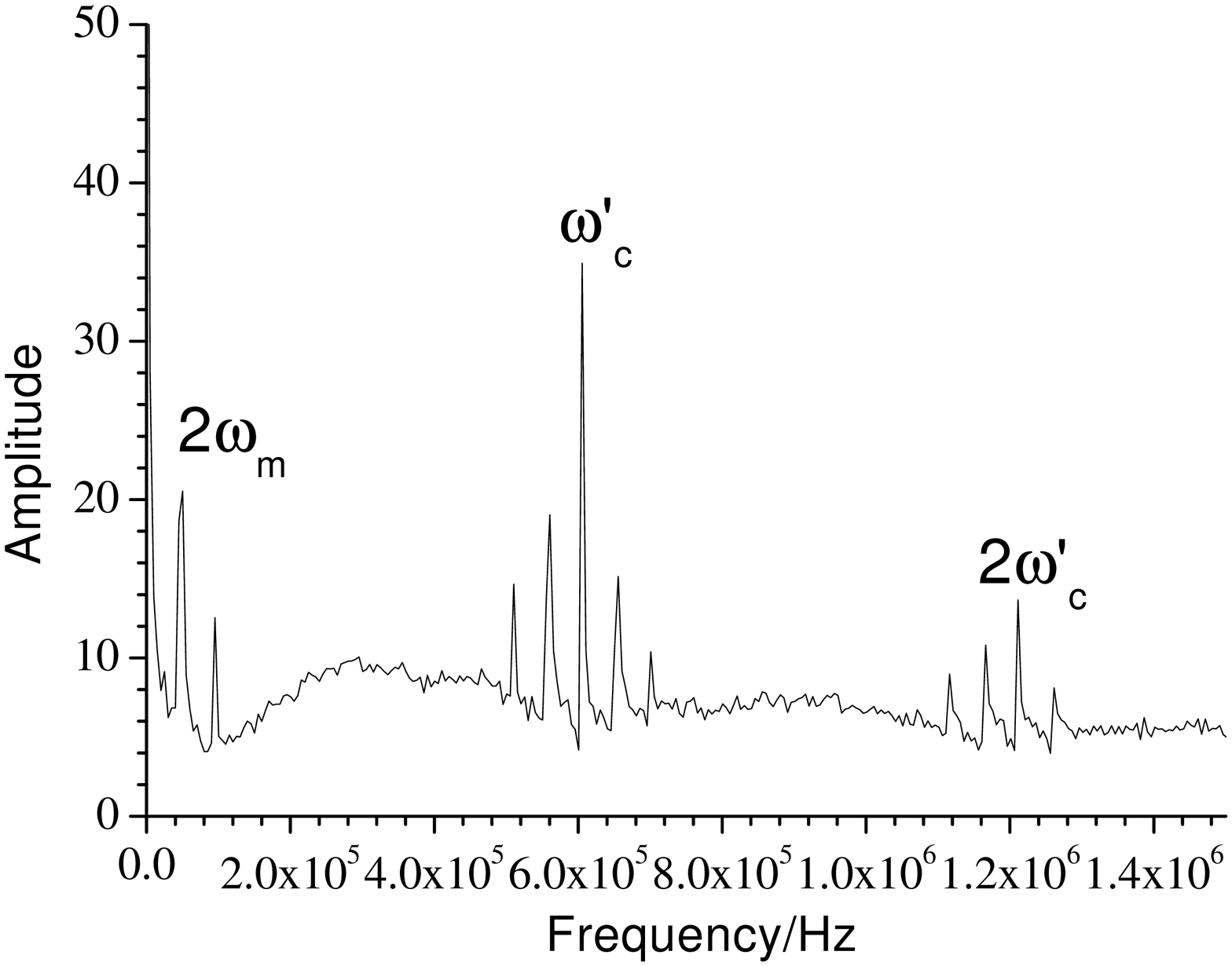}
	\end{center}
  \caption{\label{fig6} Observation of the stable orbit of an ion cloud subjected to an axialisation drive and strong laser cooling: (a) image of the cloud showing two areas of fluorescence where the orbit takes the cloud through the laser beam; (b) photon-photon correlation plot (note that the sloping background comes mainly from background scattered light); (c) FFT of trace (b) showing modulation mainly at $2\omega_m, \omega_c^{\prime}$ and $2\omega_c^{\prime}$. (The observed weaker combination frequencies arise as an artefact of this technique and do not indicate that the ions are moving at these frequencies.)}
\end{figure}

When a relatively large cloud of ions was loaded into the trap, one interesting phenomenon that was observed was the production of large stable orbits, when the damping is strong (case (c) in section 2). This is illustrated in Figure \ref{fig6}a. A large circular orbit results in two areas of fluorescence, where the orbit interacts with the laser beam. The observation direction in our experiment is perpendicular to both the trap axis and the laser beam. The ICCD images alone cannot distinguish between a single cloud moving coherently with a large magnetron radius (i.e., a large amplitude motion of the centre of mass of the cloud) and a rotating ring of ions. In an attempt to resolve this issue, photon-photon correlation was used. In this technique, consecutive detected photons are used to start and stop the TAC and the 
resulting \lq\lq waiting times" are collected in the MCA. As the motion takes the ions out of 
the laser beam altogether, the magnetron motion will appear as a strong modulation of the resonance fluorescence (Figure \ref{fig6}b). The motion at the modified cyclotron frequency 
also modulates the fluorescence, as a result of the Doppler effect taking the ion in
and out of resonance with the laser. Taking a fast Fourier transform (FFT) of the spectrum of time intervals between photons allows information about both ion oscillation frequencies to be extracted \cite{photon}.

Normally there is no correlation between photons emitted by different ions, because of the random phase difference between the motions of the ions. Therefore the correlation trace consists mainly of a smooth background in the form of an exponential decay with a time constant determined by the overall count rate. Any correlation peaks have a maximum fractional area of $1/N$, where $N$ is the number of ions. However, if the ions are all moving coherently (i.e., with fixed relative phases), a large modulation will be seen as all photons are now correlated. If the ions move together in a single cloud, the modulation will be mainly at twice 
the magnetron frequency, as the motion takes the ions through the laser beam twice each orbit. However, $N$ ions moving in a ring will give modulation at $N$ or $2N$ times the magnetron frequency. The FFT of our data (Figure \ref{fig6}c) shows clear modulation at twice the magnetron frequency and at the modified cyclotron frequency, showing that the cloud moves as 
a whole. If the ion cloud were not moving as a whole, then photons from different ions would be uncorrelated, so the modulation depth would be small. If the ions were in a ring configuration, then there would be several correlation peaks per magnetron period, which would give rise to high harmonics of the magnetron frequency in the FFT plot. Note that the background of the correlation plot shown in Figure \ref{fig6}b is mainly due to scattered light, and not to fluorescence photons from the ions. This is because the measured fluorescence rate is very low when the orbit size is large.

\section{Conclusions}

We have demonstrated that axialisation can be used in conjunction with laser cooling to give very tight confinement of a single ion to the axis of a Penning trap. The axialisation technique is complementary to the rotating wall technique, which is very effective for large ion plasmas but is not applicable for small clouds of ions and especially for a single ion. We have also shown that under certain conditions a cloud of ions in the trap will move coherently in an orbit having a large magnetron radius. This occurs when laser cooling is combined with a strong axialisation drive. The laser cooling rate for the magnetron motion was measured for a small cloud of ions and was found to be enhanced by the axialisation process. This should allow a single ion in a Penning trap to be cooled closer to the Lamb-Dicke regime than is
otherwise possible in this type of trap. The small size of the image of a single axialised ion indicates that the final temperature is much lower than that obtained without axialisation. As a result, single ions in a Penning trap may be suitable for applications in quantum information processing; it should also be possible to prepare small strings of a few ions along the magnetic field axis of a Penning trap. Although in some respects the presence of a magnetic field is a disadvantage, there are proposals for experiments for quantum information processing
where the presence of a magnetic field is required \cite{wunder}. Furthermore, decoherence rates for trapped ions may be lower in the Penning trap than they are in the rf trap due to the larger size of the electrodes and the absence of the rf confining potential. In the situation where large numbers of ions are used for quantum computing, the Penning trap also has the advantage that a two-dimensional array of ions can be formed and imaged \cite{boll}.

\ack We gratefully acknowledge the support of the EPSRC and the European Union (QUBITS project of the QIPC Programme) for this work.

\end{document}